# Body sway responses to pseudorandom support surface translations of vestibular loss subjects resemble those of vestibular able subjects

**V. Lippi[1], L. Assländer[2], E. Akcay[3], and T. Mergner[4]**

[1]Control Systems Group, Technische Universität Berlin, Germany; [2]Sensorimotor Performance Lab, University of Konstanz, Germany; [3]Kocaeli University, Kocaeli, Turkey; [4]Neurological University Clinics, Freiburg, Germany

## A B S T R A C T

Body sway responses evoked by a horizontal acceleration of a level and firm support surface are particular in that the vestibular information on body-space angle BS resembles the proprioceptive information on body-foot angle BF. We compared corresponding eyes-closed responses of vestibular-able (VA) and vestibular-loss (VL) subjects, postulating a close correspondence. In contradistinction to previous studies, we used an unpredictable (pseudorandom) stimulus and found that the eyes-closed and eyes-open responses of the VA closely resembled those of the VL subjects, as expected. We further conclude that the vestibular signals coding head linear translation in VA subjects has in this case too little functional relevance to cause a notable difference between the subject groups.

## 1. Introduction

Stance perturbations evoked by support surface translational acceleration, often used in postural control studies, differ in several respects from those evoked by support surface tilt. Both stimuli evoke body sway responses around the ankle joints, but they differ considerably in their biomechanical impact and physiological response mechanisms. For example, while body-space angle (BS) and body-foot angle (BF) differ during support surface tilts, they are mechanically coupled during support surface translations. Correspondingly, the support surface represents a space reference for the proprioceptive system during translations, but not during tilts. Hence body position can be calculated equivalently from proprioceptive and vestibular sensory inputs during translations (Fig. 1A). In this study, we asked how sway responses to support surface translations of subjects with intact vestibular function (vestibular able, VA, subjects) compare to those of subjects without vestibular function (vestibular loss; VL subjects). The additional vestibular input available to VA subjects consists of a head rotation signal that is redundant to the proprioceptive input and a linear head acceleration signal.

However, this vestibular linear acceleration signal appears to be of little relevance in postural feedback control, since it may otherwise endanger control stability [1].

Since vestibular head rotation signals are redundant and vestibular head translational acceleration signals probably do not benefit balance during support surface translations, we hypothesize that VL subjects show hardly any difference in sway responses as compared to VA subjects.

Several previous studies recorded sway responses to support surface translation and compared the responses of VA versus VL subject groups using sine-wave stimuli. They showed that subjects moved with the platform translation at low stimulus frequencies (<0.5 Hz), maintaining an upright posture (“riding the platform” response pattern) and tended to maintain head and trunk stationary in space while the leg segment moved with the platform at higher frequencies [2–4]. This behavior was observed for eyes-open and eyes-closed conditions. As pointed out by Dietz et al. [2], and also [5], however, prediction may have shaped the response patterns since these were obtained with periodic (sinusoidal) stimulation. To avoid this prediction, the latter authors [5] used a pseudo-random stimulus pattern when characterizing the effects of vision, perturbation amplitude, and perturbation frequency in VA subjects. An interesting conclusion from the work of the latter authors [5] was that, despite considerable individual differences in response, the subjects showed similarities in the balancing of their center of mass (COM) as the controlled variable.

In our study, we compared body sway responses to support surface translations between VA and VL subjects. We used a pseudo-random stimulus pattern for the support surface translation stimuli to avoid predictive contributions and compared frequency response functions (FRF) for eyes-open (EO) as well as eyes-closed (EC) conditions.

## 2. Methods

Seven VL subjects (3 males and 4 females) and seven VA subjects (5 males, 2 females) of similar age per group (means 19 years and 23 years, respectively) participated in this study. VL subjects were recruited from a school for hearing loss based on two criteria: hearing loss due to meningitis which was treated with ototoxic medicine and the inability to stand on foam rubber with eyes closed. Recruited subjects were then clinically tested for vestibular-ocular-reflexes using a rotation chair and caloric stimulation. Subjects included in this study did not show nystagmus in response to those stimuli. The study was approved by the Ethics Committee of the Freiburg University Clinics and was in accordance with the 1964 Helsinki Declaration in its latest revision.

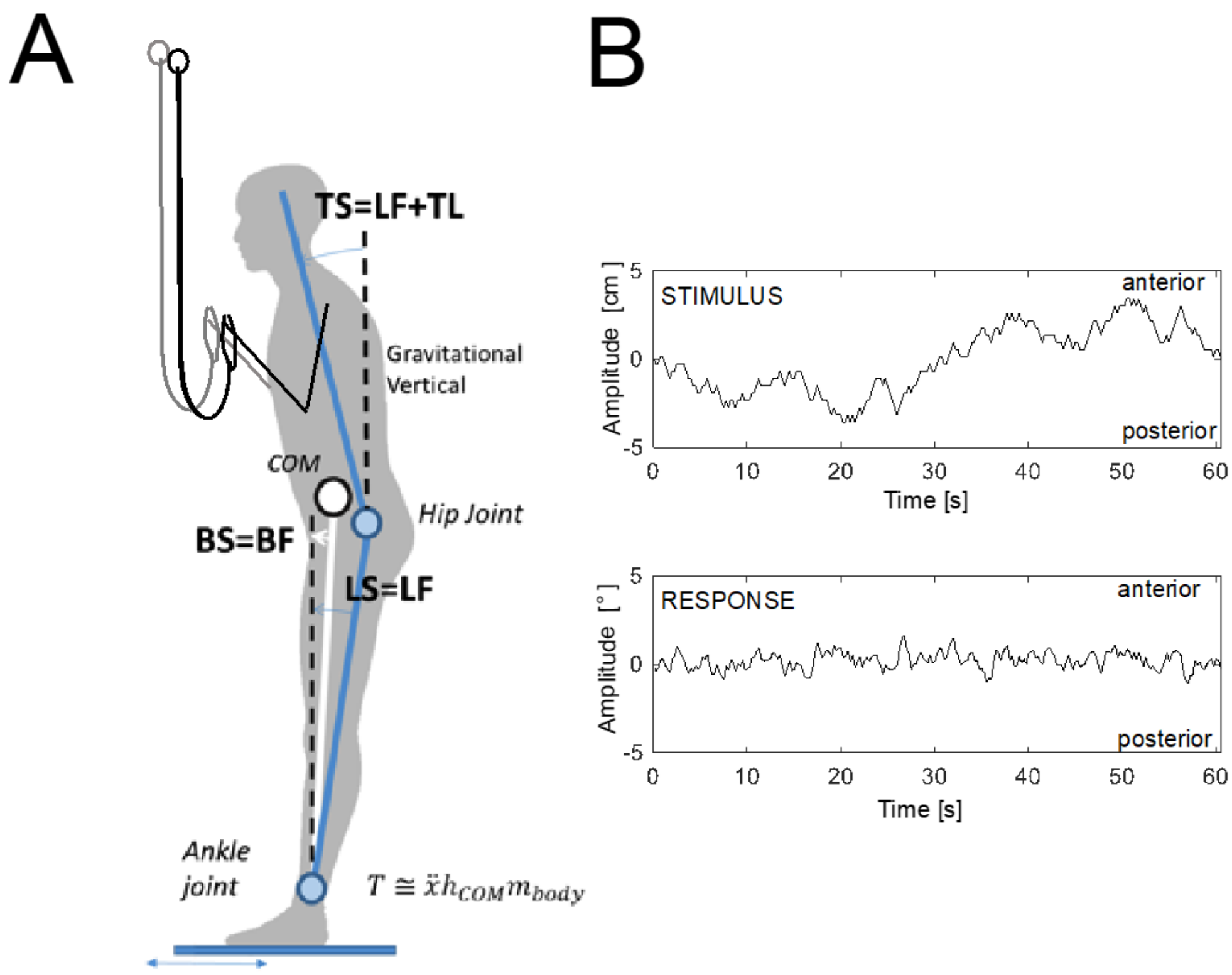

Fig 1. A. Subject standing on a translating surface, double inverted pendulum mechanics and definition of the observed kinematic variables. BS = body in space represents the angular position of the body COM with respect to the gravitational vertical passing through the ankle joint. BF = body-foot represents the angle of the COM with respect to the support surface. In the considered scenario BF=BS because the support surface is horizontal. The same notation is used for the sway of the body segments: *LS = leg in space*, *TS = trunk in space*. Joint angles are addressed as *LF = leg to foot* and *TL = trunk to leg*. The support surface translation produces an inertial force on the body, which is proportional to its acceleration $\ddot{x}$. Subject held handles with ropes fixed on the ceiling, which provide stability with stretched arms, but no spatial orientation cues with arms bent during experiments. B. PRTS stimulus with peak-to-peak amplitude of 7 cm (upper panel) and body COM sway of a subject during one stimulus.

Subjects were standing upright on a motion platform [1,6] that was moving in anterior-posterior direction. Body sway responses were measured with markers attached at hip and shoulder levels using an optoelectronic device (Optotrak 3020; Waterloo, Canada). A PC with custom-made programs was used to generate the support translation stimuli. Another PC was used to record the stimuli and the body sway from body-marker positions at a sampling frequency of 100 Hz using software written in LabView (National Instruments; Austin, USA). These measures were then exported to Matlab (The Mathworks, Natick, USA) for further analysis. Stimuli were constructed from pseudo-random-ternary sequences (PRTS, see [7] and Fig. 1B). The 60.5s-long sequence was repeated 6 times to obtain the final stimulus. The stimulus has a broad frequency range (0.01-2 Hz) and was applied at peak-to-peak amplitudes of 3.5 cm and 7.0 cm with eyes closed (EC) and with eyes open (EO). Each condition was tested twice.

Subjects held the handle of a safety rope in each hand (Fig. 1A), which gave no support or spatial orientation cues with flexed arms during trials, but would support the body when the arms were stretched. VL and VA subjects also wore straps fixed to the trunk with the same function as the safety ropes. Furthermore, in the experiments of the VL subjects, one of the experimenters would stand next to the VL subject ready to help to stabilize (which was never necessary). Subjects listened to non-rhythmic audio books to draw subjects' attention away from the balancing task so that they would perform this task mainly in an automatic way and to minimize auditory orientation cues. In the trial with visual input, the visual field was fully illuminated, unrestricted and comprised approximately half of the laboratory (distance to the front wall of 4 m), with a rich visual scenario providing a broad range of contrasts and spatial frequencies as well as 3D structures.

Hip and shoulder marker movements were used to calculate trunk and leg segment angles in space, as well as body COM sway (BS) using anthropometric tables [8]. Sway responses were averaged across all PRTS sequence repetitions within subjects, where the first cycle of each trial was always discarded to avoid transient responses. Spectra of the stimulus and all outcome measures were calculated using the Matlab function "fft", averaged across sequence repetitions, and frequency response functions of the angular response to the translation stimulus (°/cm) were calculated thereof [6]. Finally, frequency response functions were expressed as gain and phase. Coherence was calculated as the squared cross-power spectrum $G_{xy}(f)$ divided by the product of sway response $G_{xx}(f)$ and stimulus power spectra $G_{yy}(f)$, *as given by:*

$$C_{xy} = \frac{\left|G_{xy}(f)\right|^2}{G_{xx}(f)G_{yy}(f)}$$

The FRF graphs shown below give mean and standard deviation of gain and phase across subjects. Gain is the amplitude ratio of response and stimulus, where for example gain values of 0, 1, and 2 would represent the ratio of 0°, 1°, and 2° body angular sway evoked by a 1 cm translation of the platform at a given frequency. Phase gives the relative timing of sway response and stimulus. Coherence can be interpreted as a measure of sway response to random sway not evoked by the stimulus. A coherence value of 1 would indicate that there is no random sway at a given frequency, while the random sway component is larger for smaller coherence values.

Frequency response functions (FRFs) were statistically compared across groups and conditions using bootstrap hypothesis tests [9]. The norm of the difference between the average FRFs of two conditions was calculated ($\hat{\vartheta}$). In a second step, an equal number of cycles as contained in the original data-set was drawn with replacement. The norm of the difference was calculated from this new bootstrap data set ($\hat{\vartheta}^*$). The statistic was then calculated using

$$\hat{t}^* = \frac{\hat{\vartheta}^* - \hat{\vartheta}}{\hat{\sigma}^*}$$

where the variance (sigma) was obtained from nested bootstrap tests using 200 samples. The procedure was repeated 10000 times and the results sorted in descending order. The tests show a significant difference, if the norm of the difference between the two FRFs (Theta) is larger than 95% of the bootstrap statistic (see examples in Fig. 4,Aa and Ba). To test the performance across frequencies, we calculated the same statistics for each frequency point separately (see Fig. 4Ab and Bb). All statistical results are provided in the supplementary material, while the main results are described in the text.

## 3. Results

### *3.1 Eyes-closed condition*

*The sway of body COM in space.* With the 3.5 cm translations EC (Fig. 2A*, BS*), BS gain was close to zero in the low-frequency range (<0.11 Hz), similarly in both VA and VL subjects. The difference between the averages of the FRFs for VA and VL was statistically not significant ($p > 0.05$). Phase showed large variability in this range, and coherence was low (<0.5). The larger standard deviation in phase does not correspond to an actual large variability in the FRF. The reason is that the FRF is defined in the complex domain. When the gain is zero, the phase is not defined, and as the gain approaches zero, the phase is subject to large variations due to noise that is associated to small variations of the complex value.

In the mid-frequency range (0.11-0.16 Hz), gain increased to values around unity, coherence increased to values somewhat below unity, and phase developed a lag that passed through 180°. In the high-frequency range (0.17-2.19 Hz) gain, phase, and coherence leveled off. Body sway (BS) responses to the 7.0 cm translation stimuli (Fig. 2B, *BS*) were very similar across both subject groups, but BS FRFs were significantly different between VL and VA with both translation amplitudes during eyes-closed conditions ($p<0.05$). The difference was mainly due to the response at mid frequencies. The findings were similar with both the 3.5 cm and 7.0 cm stimuli ($p>0.05$).

*Sway of the trunk segment in space* (Fig. 2A*, TS*). Gain, phase, and coherence curves generally resembled those obtained for the leg segment and showed a qualitatively similar behavior at both stimulus amplitudes and for VL and VA subjects (e.g. the difference was here statistically not significant, $p>0.05$). However, TS gain values in the mid- to high-frequency range (0.17 Hz and 2.19 Hz) was considerably larger in VA subjects as compared to VL. This difference between the two groups was significant both with amplitude 7.0 cm ($p<0.05$) and with amplitude 3.5 cm ($p<0.05$). Thus, VA subjects allowed for higher trunk sway amplitudes as compared to VL subjects, although with a similar relative temporal coupling with the leg segment (a bootstrap test on the phase profile did not show significant difference between the two groups, $p>0.05$). With EC, there was no significant difference between the results obtained with the 3.5 cm and 7 cm amplitude stimuli ($p>0.05$).

*Sway of the leg segment in space* (Fig. 2A*, LS*). The results closely match those of the BS responses described above. This is not surprising since the body COM is located not far above the upper end of the leg segment (compare [8]).

### *3.2 Eyes-open condition*

With eyes-open (EO), the largest difference compared to the EC condition was observed in the low-frequency range. Here, the sway responses showed gain values that increased from about 0.1 °/cm at 0.017 Hz to about 0.4°/cm at 0.11 Hz, with less variability in the phase and coherence values closer to unity (Fig. 3A, B). EC and EO responses were significantly different ($p<0.05$). In contrast to the almost absent response with EC in the low-frequency range, the gain rises here with increasing frequency to reach roughly similar values as with eyes closed in the mid and high-frequency ranges. The results of VA and VL subjects again resembled each other closely (corresponding BS FRFs were not significantly different, $p>0.05$). An exception was the gain of TS responses, which were slightly larger in VA subjects than in VL subjects (Fig. 3A, B; $p<0.05$ for both the stimulus amplitudes). Generally, gain variability tended to be smaller with EO as compared to EC (compare to Figs. 2A, B). The amplitude of the translation stimulus produced only in the case of VL subjects with EO a significant difference in the response ($p<0.05$).

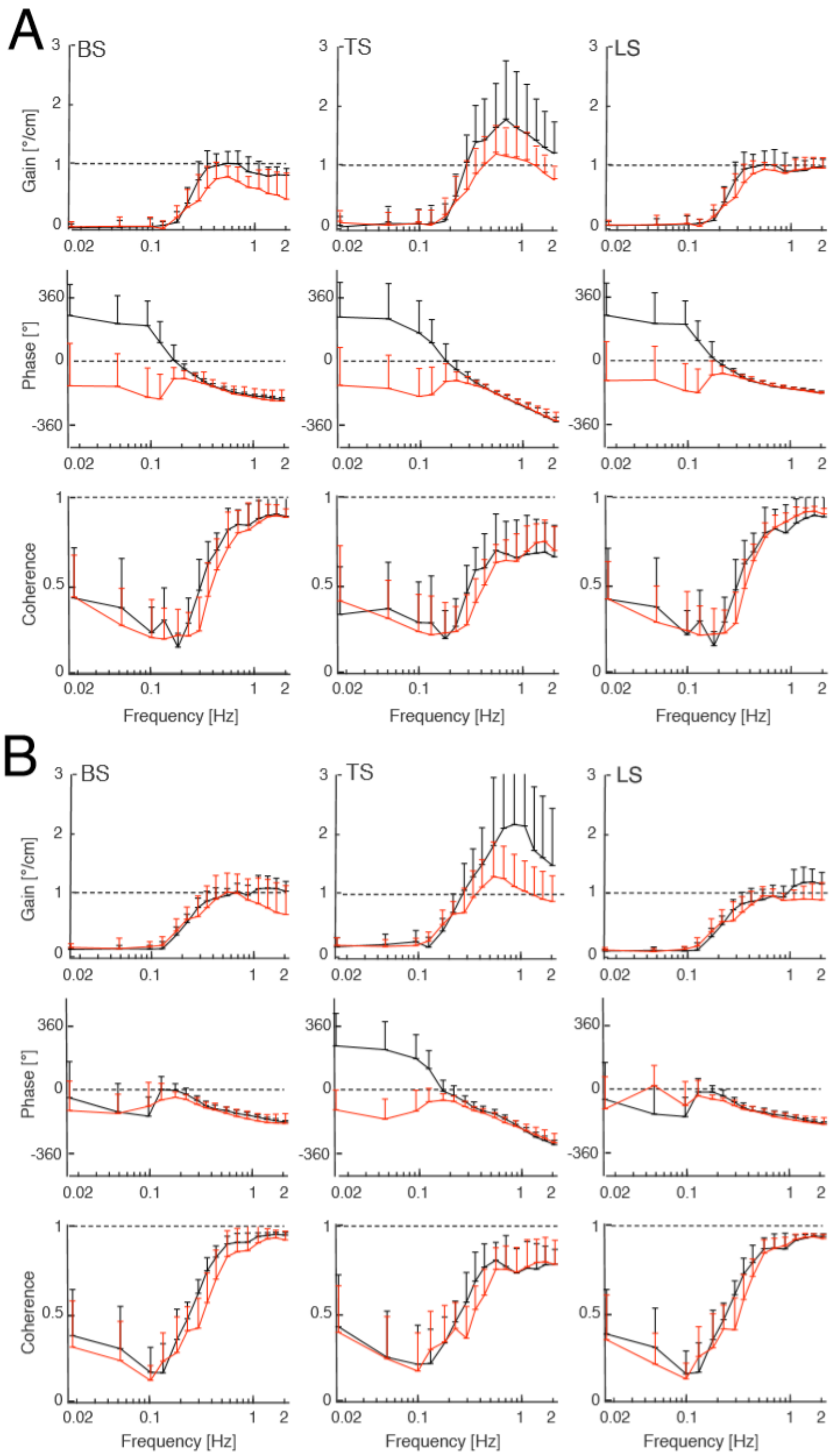

Fig. 2A,B. Eyes-closed responses. Support surface translation responses of the two subject groups (VA, vestibular able: black; VL, vestibular loss: red) with the eyes closed for pp (peak-to-peak) amplitude 3.5 cm (A) and pp 7 cm (B) support surface displacement in the body's sagittal plane. Plotted are angular sway excursions of the body COM in space (BS), of the trunk in space (TS) and of the legs segment in space (LS) about the ankle joints in terms of mean and +S.D. values (7 subjects in each group).

## 4. Discussion

We hypothesized that sway responses to support surface translation are similar in VL subjects compared to VA subjects. This primarily relates to the EC condition and less so to the EO condition, since vision may level out to a large extent the vestibular loss in the VL subjects. Our experiments confirmed our hypothesis for the control of the body COM and the leg segment, which are the most relevant for successful balancing. As an exception, with eyes closed, VA subjects had larger TS sway as compared to VL subjects at frequencies above 0.3 Hz. However, this difference had only a relatively small effect on LS and BS. A tilt of the leg segment (LS) makes a 3.5 times larger contribution to the body COM tilt (BS) as compared to the trunk segment (TS). The main reason is that the trunk (about 2/3 of body weight) is translated on the top of the leg segment during a leg segment tilt. Such a small effect of TS on BS is visible at the smaller stimulus amplitude (Fig. 2A) and at frequencies above 1 Hz for the larger stimulus amplitude (Fig. 2B). However, this difference in contribution does not explain why the TS differences between VL and VA, e.g. at around 0.5 Hz, are hardly reflected in BS. Here, the difference in phase between LS and TS leads to a partial cancellation of the amplitudes. Both in EO and EC conditions, TS exhibits a larger phase lag compared with LS, which is especially evident at higher frequencies (see Fig 2A, B and Fig. 3A, B). We argue that the difference between these sway responses is so small during eyes-closed conditions because humans tend to rely strongly on the proprioceptive reference to the support surface during the translational perturbations. One reason in VA subjects could be that the proprioceptive reference contains less sensory noise than the corresponding vestibular one [10]. Interestingly, evidence showing that proprioceptive input suffices to explain trunk stabilization in the body sagittal plane has been reported in [11].

Previous work studying the role of the vestibular system for the stabilization of standing posture during sinusoidal support surface translations reported for subjects with well compensated vestibular loss functions similar results as for normal controls during eyes closed conditions [12]. Thus, predictability inherent in the sinusoidal stimuli appears not to be the decisive factor of these previous results, as the similarity mostly holds for the present study in which the translation stimuli were unpredictable.

The finding that with EC, both subject groups showed hardly any sway in the low-frequency range, but mainly in the mid- and high-frequency ranges, appears surprising and requires further experimental consideration in the future. This includes the question of why sway in the low-frequency range was larger in EO than in EC, since intuitively, one may expect better postural stabilization (i.e. lower sway response gain) when visual information is available. The obtained EC and EO responses at low frequencies are in agreement with earlier studies observing a "ride of the platform" during eyes closed and a movement of the leg segment underneath the more stable body center of mass during eyes open. Using virtually unpredictable stimuli, our data thus confirms earlier findings of experiments using predictable sinusoidal translations [2-4,12]. The present study and the previous work does not yet, however, provide a formal description of the underlying sensorimotor control mechanisms.

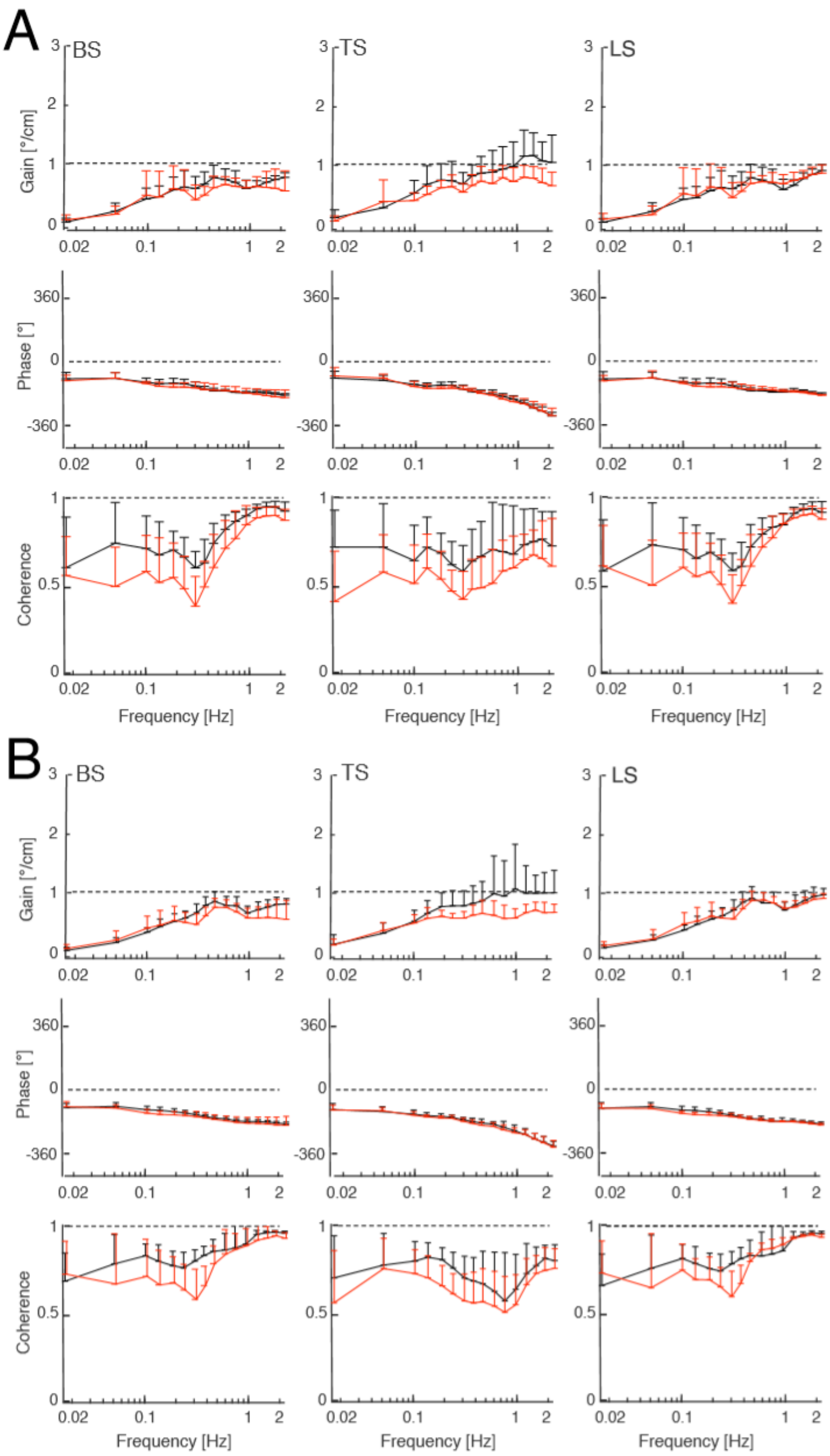

Fig. 3A, B. Eyes-open responses. Support surface translation responses of the two subject groups (VA, vestibular able: black; VL, vestibular loss: red) with the eyes closed for peak to-peak (pp) amplitude 3.5 cm stimuli (A) and pp 7 cm support surface displacement in the body's sagittal plane (B). Plotted are angular sway excursions of the body COM about the ankle joints (BS), the sway of the trunk (TS) and the sway of the legs (LS) in terms of mean and SD values (7 subjects in each group).

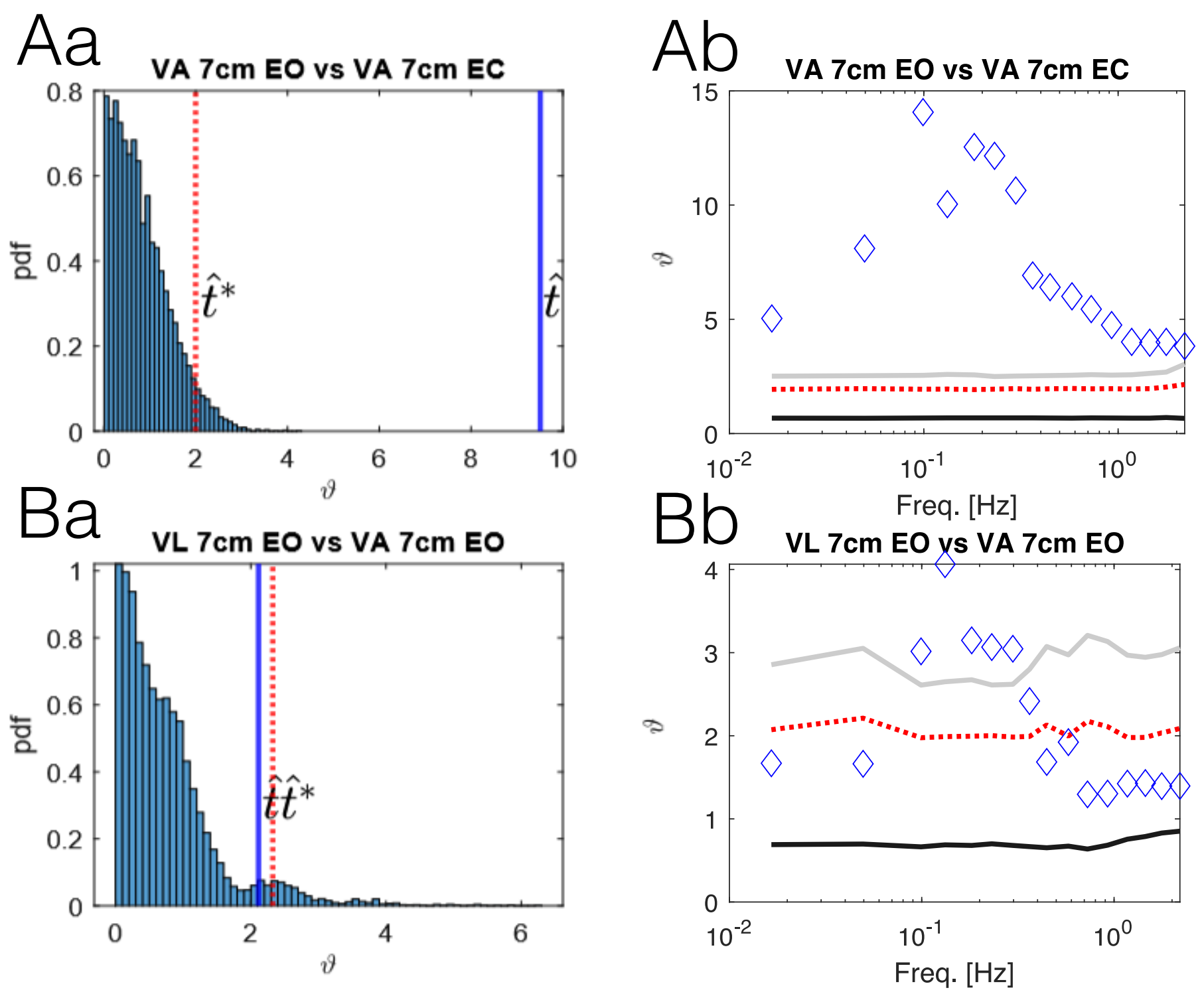

Fig. 4A,B. Two example sets comparing between BS FRFs (heading on top in Aa,Ab Vestibular able subjects stimulated with pp 7 cm translation, eyes open versus eyes closed, in Ba,Bb. Vestibular loss subjects compared to vestibular able subjects for 7 cm stimulus with eyes open). Left Columns (Aa,Ba): Blue histograms (Aa, Ba) show the distributions of the statistics (pdf = probability density function). The dotted red line labelled $\hat{t}^*$ represents the *p=0.05* confidence interval, the blue solid line labelled $\hat{t}$ represents the statistics computed on the dataset, the null hypothesis: i.e. the hypothesis that the average of the two groups is the same is rejected if $\hat{t} < \hat{t}^*$. if the blue line is on the right side with respect to the red line. Right Columns (Ab,Bb): Frequency-wise comparison (16 tests for the 16 frequencies) of the FRFs (same data sets as in Aa and Ba). The dotted red lines represent the *p=0.05* confidence interval, the blue diamonds represent the $\hat{t}$ for each frequency computed on the dataset. The null hypothesis (FRFs difference is zero at the considered frequency) is rejected if $p < 0.05$, i.e. the blue diamond is above the red line. The light grey line represents the *p=0.01* confidence interval; the black line represents the median of the statistics following the null hypothesis computed with the bootstrap method. Note different scaling of ordinate in Ab and Bb.

## 5. Conclusions

The comparison of the body sway induced by support surface translations in VL and VA subjects showed overall little difference between the two groups and between stimulus amplitudes. Visual conditions, conversely, had a strong effect on the response producing two specific response patterns (shown in Fig. 2 and Fig. 3, respectively). Body sway responses did not suggest the presence of a non-linear gain behavior when changing stimulus amplitudes (panels A and B closely resemble each other in EC, Fig. 2 and EO, Fig. 3), contrasting with what has been observed for support surface tilt [6]. This result suggests that the relevant part of the observed behavior is here mainly due to the effect of linear control systems, such as a mainly proprioceptive compensation of the body lean position produced by the gravitational torque from the induced body sway, and to the effect of passive stiffness. This is consistent with the observation that VA and VL subjects behave similarly, as the aforementioned posture control mechanisms here do not require the vestibular input.

**Author contribution**

All four authors were responsible for study design, data collection, analysis and manuscript writing.

## References

[1] T. Mergner, A neurological view on reactive human stance control, Annu. Rev. Control. 34 (2010) 177–198. https://doi.org/10.1016/j.arcontrol.2010.08.001.

[2] V. Dietz, M. Trippel, I.K. Ibrahim, W. Berger, Human stance on a sinusoidally translating platform: balance control by feedforward and feedback mechanisms, Exp. Brain Res. (1993). https://doi.org/10.1007/BF00228405.

[3] S. Corna, J. Tarantola, A. Nardone, A. Giordano, M. Schieppati, Standing on a continuously moving platform: Is body inertia counteracted or exploited?, Exp. Brain Res. (1999). https://doi.org/10.1007/s002210050630.

[4] G.N. Gantchev, S. Dunev, N. Draganova, On the problem of the induced oscillations of the body., Agressologie. (1972).

[5] D. Joseph Jilk, S.A. Safavynia, L.H. Ting, Contribution of vision to postural behaviors during continuous support-surface translations, Exp. Brain Res. (2014). https://doi.org/10.1007/s00221-013-3729-4.

[6] G. Hettich, L. Assländer, A. Gollhofer, T. Mergner, Human hip-ankle coordination emerging from multisensory feedback control, Hum. Mov. Sci. (2014). https://doi.org/10.1016/j.humov.2014.07.004.

[7] R.J. Peterka, Sensorimotor integration in human postural control, J. Neurophysiol. (2002). https://doi.org/10.1152/jn.2002.88.3.1097.

[8] D.A. Winter, Biomechanics and Motor Control of Human Movement: Fourth Edition, 2009. https://doi.org/10.1002/9780470549148.

[9] A.M. Zoubir, B. Boashash, The bootstrap and its application in signal processing, IEEE Signal Process. Mag. (1998). https://doi.org/10.1109/79.647043.

[10] H. Van Der Kooij, R.J. Peterka, Non-linear stimulus-response behavior of the human stance control system is predicted by optimization of a system with sensory and motor noise, J. Comput. Neurosci. (2011). https://doi.org/10.1007/s10827-010-0291-y.

[11] J.H. van Dieën, P. van Drunen, R. Happee, Sensory contributions to stabilization of trunk posture in the sagittal plane. *J. Biomechanics* (2018), https://doi.org/10.1016/j.jbiomech.2017.07.016

[12] J.J. Buchanan, F.B. Horak, Emergence of postural patterns as a function of vision and translation frequency, J. Neurophysiol. (1999). https://doi.org/10.1152/jn.1999.81.5.2325.